\documentclass[prd,aps,floats,preprintnumbers,twocolumn]{revtex4}
\usepackage{graphicx, epsfig}
\usepackage{amsmath}
\usepackage[T1]{fontenc}
\usepackage[latin1]{inputenc}
\usepackage{graphicx}
\usepackage{color}
\newcommand{\beq}{\begin{equation}}
\newcommand{\eeq}{\end{equation}}
\newcommand{\bea}{\begin{eqnarray}}
\newcommand{\eea}{\end{eqnarray}}

\newcommand{\aap}{{Astron.~Astrophys.}}

\newcommand{\mnras}{{Mon.~Not.~R.~Astron.~Soc.}}

\def\gs{\mathrel{\raise1.16pt\hbox{$>$}\kern-7.0pt %
\lower3.06pt\hbox{{$\scriptstyle \sim$}}}}         %
\def\ls{\mathrel{\raise1.16pt\hbox{$<$}\kern-7.0pt %
\lower3.06pt\hbox{{$\scriptstyle \sim$}}}}         %

\begin{document}

\setlength{\unitlength}{1mm}
\title{Constraints on primordial non-Gaussianity from WMAP7\\ and Luminous Red Galaxies power spectrum and forecast for future surveys}
\author{Francesco De Bernardis$^{1,2}$, Paolo Serra$^{2}$, Asantha Cooray$^{2}$, Alessandro Melchiorri$^{1}$}
\affiliation{$^1$Physics Department and INFN, Universita' di Roma ``La Sapienza'', Ple Aldo Moro 2, 00185, Rome, Italy.}
\affiliation{$^2$Center for Cosmology, Dept. of Physics \& Astronomy, University of California Irvine, Irvine, CA 92697.}


\bigskip

\begin{abstract}
We place new constraints on the primordial local non-Gaussianity
parameter $f_{NL}$ using recent Cosmic Microwave Background anisotropy and galaxy clustering data.
We model the galaxy power spectrum according to the halo model, accounting for a scale dependent bias correction
proportional to $f_{NL}/k^2$. We first constrain $f_{NL}$ in a full $13$ parameters analysis that includes $5$ parameters of the
halo model and $7$ cosmological parameters. Using the WMAP7 CMB data and the SDSS DR4 galaxy power spectrum,
we find $f_{NL}=171^{+140}_{-139}$ at $68\%$ C.L. and
$-69<f_{NL}<+492$ at $95\%$ C.L.. We discuss the degeneracies between $f_{NL}$ and other cosmological parameters. Including SN-Ia data
and priors on $H_0$ from Hubble Space Telescope observations we find a stronger bound: $-35<f_{NL}<+479$ at $95\%$.
We also fit the more recent SDSS DR7 halo power spectrum data finding, for a $\Lambda$CDM+$f_{NL}$ model,
$f_{NL}=-93\pm128$ at $68\%$ C.L. and $-327<f_{NL}<+177$ at $95\%$
C.L.. We finally forecast the constraints on $f_{NL}$
from future surveys as EUCLID and from CMB missions as Planck
showing that their combined analysis could detect $f_{NL}\sim 5$.
\end{abstract}


\maketitle

\section{Introduction}

The standard paradigm of structure formation relies on the inflation
\cite{starob, albrecht}. As shown in \cite{M,Bardeen1983},
quantum-mechanical fluctuations in the scalar field driving
inflation lead to primordial density perturbations responsible of
the large scale structures we observe today. Although the simplest
assumption is that these fluctuations were Gaussian distributed
\cite{bardeen}, there are several inflationary models
\cite{linde2}-\cite{sasaki} involving the existence of a primordial
non-Gaussianity. A detection or exclusion of non-Gaussianities would
hence be of fundamental interest for the understanding the physics
of the primordial Universe. Several cosmological observables and
methods can be used to constrain non-gaussianities. \\Cosmic
Microwave Background (CMB) anisotropies provide the most direct
method for the detection of primordial non-Gaussianity (see e.g.
\cite{Bartolo:2010qu}) through, for example, measurements of the
three-point correlation function \cite{Matarrese:1997sk} (or
equivalently the bispectrum) which is non-zero in presence of
non-Gaussianities. The large scale structure of the Universe is also
affected by non-Gaussianities that may be detected by looking at the
bispectrum or the trispectrum of galaxy distribution. The abundance
of galaxy clusters, that depends on the tails of the density
probability distribution, is also sensitive to any deviation from
gaussianity. Non-Gaussianity has a direct impact on the clustering
of dark matter halos by changing their mass and correlation function
(\cite{lu_mat}-\cite{Dalal:2007cu}).\\ A common way to parameterize
primordial non-Gaussianities is to introduce a quadratic correction
to the potential \cite{Gangui:1993tt}\cite{Komatsu:2000vy}:
\begin{equation}\label{fnl}
\Phi=\phi+f_{NL}(\phi^2-\langle\phi^2\rangle)
\end{equation}
where $\Phi$ is the primordial potential and $\phi$ is a gaussian random field. In this case the non-Gaussianity is a local type correction whose amplitude is given by $f_{NL}$.
The most recent constraint on $f_{NL}$ from CMB gives $-10<f_{NL}<+74$ at $95\%$ C.L. from bispectrum analysis of WMAP-7 years data \cite{Komatsu:2010fb}, improving the WMAP5 constrain ($-9<f_{NL}<+111$) \cite{Komatsu:2008hk}. The authors of \cite{Calabrese:2009bu} used a different estimator applied to WMAP5 data, finding $f_{NL}=-13\pm62$ at $68\%$ C.L., while \cite{Rudjord:2009mh}, using the needlet bispectrum applied to the same data, found $f_{NL}=+84\pm40$ at $68\%$ C.L..\\In \cite{Verde:1999ij}-\cite{Reid:2010vc}-\cite{Dalal:2007cu}-\cite{Matarrese:2008nc}-\cite{Slosar:2008hx} it has been shown that a quadratic correction to the potential like that of equation (\ref{fnl}) produces a scale dependence in the bias of the galaxy clustering with respect to matter distribution. In particular, a scale dependent term $\Delta b(k)$ arises in the halo bias on larger scales (smaller $k$) and is proportional to $f_{NL}$ ($\Delta b(k)\propto f_{NL}/k^2$) hence with galaxies being more (less) clustered for positive (negative) values of $f_{NL}$. The authors of \cite{Slosar:2008hx} analyzed the galaxy power spectrum of luminous red galaxies (LRG) of the Sloan Digital Sky Survey (SDSS) to constrain this scale dependence of the bias, putting the constraint $-21<f_{NL}<+209$ at $95\%$ C.L.. In the same work other large scale datasets have been used to constrain non-gaussianity (Quasars, integrated Sachs-Wolfe effect data and photometric LRG sample) finding $-29<f_{NL}<+70$ at $95\%$ C.L. from the combination of all datasets. Recently, the authors of \cite{Xia:2010yu} obtained $+25<f_{NL}<+117$ at $95\%$ C.L. from the combination of WMAP-7 years data \cite{Komatsu:2010fb}, Baryonic Oscillations data from SDSS and Two-degree Field Galaxy Redshift Survey (2dFGRS) \cite{Percival:2009xn} and Supernovae distance moduli measurements \cite{Kowalski:2008ez} with auto correlation function measurements of radio sources from NRAO VLA Sky Survey \cite{condon98}, claiming a detection of non gaussianity at $\sim 3\sigma$.

In this paper we follow the methodology of \cite{Slosar:2008hx} and constrain $f_{NL}$ by looking at the scale dependence of the bias in current galaxy surveys data. We implement the calculation of galaxy and halo power spectrum using
the halo-model (see \cite{Cooray:2002dia} and section \ref{halomodel}) and we include in it the non-Gaussian scale-dependent correction to the bias.
We then place constraints on $f_{NL}$ by comparing this model to the Sloan Digital
 Sky Survey (SDSS) \cite{www.sdss.org} galaxy power spectrum data \cite{Tegmark:2006az} and to the halo power spectrum data obtained from the
luminous red galaxies (LRG) sample \cite{Reid:2009xm}. We include in the analysis the WMAP-7 years cosmic microwave background anisotropy data \cite{Komatsu:2010fb}. We also fit the LRG galaxy power spectrum data to the same model, including Hubble constant measurements from Hubble Space Telescope (HST, \cite{Riess:2009pu}) and Supernovae distance moduli measurements for the Union dataset \cite{Kowalski:2008ez}. Results are shown in section \ref{results}. We finally forecast the power of future Galaxy surveys in constraining non gaussianity by generating mock data for galaxy power spectrum using specifications of EUCLID \cite{Refregier:2006vt} survey combined with mock data from Planck \cite{:2006uk}
satellite and showing in section \ref{forecasts} that the combination of data from these experiments
could reach the precision required to detect even small non-Gaussianities.

\section{Halo-model}\label{halomodel}
In the halo model scenario (see \cite{Cooray:2002dia} for a detailed
review) all matter is contained in halos and, as a consequence, the
abundance of halos, their spatial distribution and their internal
density profiles are closely connected to the initial dark matter
fluctuation field. Under the assumption that galaxies are formed in
these halos of dark matter \cite{whiterees} is then possible to use
the halo model to calculate the statistical properties of
distribution of galaxies. To this aim, the basic quantity is the
halo occupation distribution (HOD, see  \cite{Peacock:2000qk}) that
encodes the information on how galaxies populate dark matter halos
as a function of halo-mass. The statistical information is contained
in the two-point correlation function of galaxies or equivalently
its Fourier transform, the galaxy power spectrum. It is hence
important to assess the number of pairs of galaxies in an individual
halo and the number of pairs of galaxies in separate halos. The
former can be shown to be related to the variance of the HOD,
$\sigma^2(M,z)=\langle N_{g}(N_g-1)\rangle$ while the latter is the
square of the mean halo occupation number $N(M,z)=\langle
N_{g}\rangle$. The galaxy power spectrum is then the sum of the
$1$-halo term describing pairs of objects in the same halo and of a
$2$-halo term for objects in different halos:
$P(k,z)=P_{1h}(k,z)+P_{2h}(k,z)$. The two terms can be written as:
\begin{equation}\label{1halo}
\begin{split}
P_{1h}(k,z)=\frac{1}{n^2_{gal}(z)}\times\\
\int dMn_{halo}(M,z)|u_{DM}(k,M,z)|^p\sigma^2(M,z)
\end{split}
\end{equation}
\begin{equation}\label{2halo}
\begin{split}
P_{2h}(k,z)=\frac{P_0(k,z)}{n^2_{gal}(z)}\times\\
\left[\int dM n_{halo}(M,z)N(M,z)b(M,z)u_{DM}(k,M,z)\right]^2
\end{split}
\end{equation}
where $n_{halo}$ is the halo mass function \cite{Sheth:2001dp},
$u_{DM}(k,M,z)$ is the normalized dark matter halo density profile
in Fourier space, $P_{0}(k,z)$ is the linear dark matter power
spectrum, $b(M,z)$ the linear bias parameter and $n_{gal}$ is the
mean galaxy number per unit of comoving volume:$$n_{gal}(z)=\int
dMn_{halo}(M,z)N(M,z).$$ For low occupied halos ($N(M,z)<1$) the
exponent $p$ of the density profile in the (\ref{1halo}) is equal to
$1$ while it is equal to $2$ otherwise \cite{Cooray:2002dia}. To
calculate the two terms (\ref{1halo}) and (\ref{2halo}) it is
necessary to assume a form for the HOD. We choose the
parameterization described in \cite{Abazajian:2004tn},
\cite{Zheng:2004id} where the HOD consists of two separated
contributions for central and for satellite galaxies:
\begin{equation}\label{ncen}
    \langle N_{cen}(M)\rangle=\frac{1}{2}Erfc\left[\frac{\ln (M_{min}/M)}{\sqrt{2}\sigma_{cen}}\right]
\end{equation}
\begin{equation}\label{sat}
\langle
N_{sat}(M)\rangle=\left[\frac{M-\gamma M_{min}}{M_1}\right]^{\alpha}
\end{equation}
where $M_{min}$, $M_1$ $\sigma_{cen}$, $\gamma$ and $\alpha$ are free
parameters of the model. In this description the mean occupation
number of central galaxies is modeled as a smoothed step function
above the minimum mass $M_{min}$, while satellite galaxies follow a
Poisson distribution with a mean given by a power low and a cut-off
at multiple $\gamma$ of the minimum mass. This $5$-parameters model
showed a good agreement with hydrodynamical and N-body simulations
and semi-analytic models \cite{Guzik:2002zp}, \cite{Berlind:2002rn},
\cite{Kravtsov:2003sg}.
\\For the halo
density profile $u_{DM}(k,M,z)$ we choose the shape of the Navarro,
Frenk $\&$ White profile (NFW) \cite{Navarro:1996gj}. The variance
of the HOD can be calculated as in \cite{Magliocchetti:2006qa}:
\begin{eqnarray}
  \sigma(M,z)=N(M,z),\hspace{1cm} N(M,z)>1\\
\sigma(M,z)=\beta(M)^2N(M,z),\hspace{1cm} N(M,z)<1
\end{eqnarray}
with $\beta(M,z)=\log_{10}(M/M_{min})/log_{10}(M_0/M_{min})$ and
$M_0$ is the mass at which the mean occupation number is equal to
$1$. This parameterization of $\sigma (M,z)$ has been shown to have
a good agreement with both semi-analytic models and hydrodinamical
simulations \cite{Viero:2009qm}, \cite{Berlind:2001xk}.
\\The halo mass function is given by the Press \& Schechter relation
\cite{Press&Schechter}:
\begin{equation}\label{PressSchechter}
    \frac{M^2n_{halo}(M,z)}{\bar{\rho}}\frac{dM}{M}=\nu f(\nu)\frac{d\nu}{\nu}
\end{equation}
where $\bar{\rho}$ is the background comoving density and $\nu$ is
defined as the ratio between the critical density required for
spherical collapse at redshift $z$ ($\delta_{sc}(z)$) and the
variance of the initial density fluctuation field $\sigma_0(M)$:
$\nu=\delta_{sc}^2(z)/\sigma^2_0(M)$. Here we choose the Sheth-Tormen model \cite{Sheth:1999mn}
 for the shape of $\nu f(\nu):$
\begin{equation}\label{ShethTormen}
    \nu
    f(\nu)=A(p)(1+(q\nu)^{-p})\left(\frac{q\nu}{2\pi}\right)^{1/2}\exp\left(-\frac{q\nu}{2}\right)
\end{equation}
with $p\sim0.3$, $A(p)\sim0.3222$ and $q\sim0.75$. The linear bias
$b(M,z)$ is then given by \cite{Sheth:1999mn}, \cite{Mo:1996zb}:
\begin{equation}\label{bias}
b(M,z)=1+\frac{q\nu-1}{\delta_{sc}(z)}+\frac{2p/\delta_{sc}(z)}{1+(q\nu)^p}
\end{equation}
\begin{figure}
  \begin{center}
  \includegraphics[width=230pt]{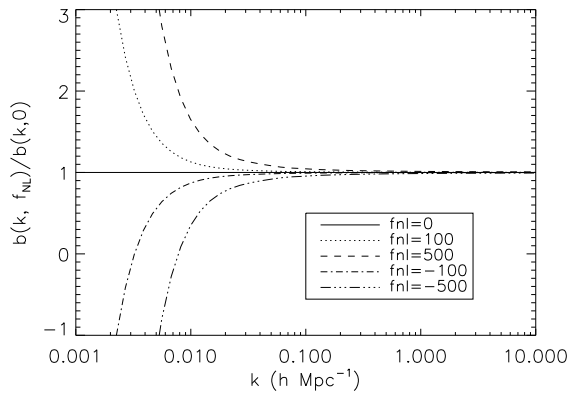}\\
  \end{center}
   \caption{Scale dependent correction to the halo bias according to the (\ref{correction}) for
   different values of $f_{NL}$.}\label{biascorrection}
\end{figure}

\subsection{Non-Gaussian corrections} The halo model described so far
allows the calculation of the galaxy power spectrum starting from
the assumption of gaussian primordial fluctuations. The existence of
deviations from gaussianity determines a correlation between small-scale
and large scale perturbations because of the quadratic correction
$f_{NL}\phi^2$ (in the case of local non-gaussianities we are
considering here) that appear in the potential
\cite{Komatsu:2001rj}, \cite{Gangui:1993tt}. As shown in
\cite{Verde:1999ij}-\cite{Matarrese:2008nc}-\cite{Slosar:2008hx}
the effect of non-Gaussian fluctuations on the galaxy power spectrum
appears on large scales through a scale dependent
correction of the halo bias. Following \cite{Slosar:2008hx} we write
this correction as:
\begin{equation}\label{deltabias}
    \Delta b(M,z,k)=\frac{3\Omega_mH_0^2}{c^2k^2T(k)G(z)}f_{NL}\frac{\partial\ln n_{halo}}{\partial\ln\sigma_8}
\end{equation}
\begin{figure}
  \begin{center}
  \includegraphics[width=230pt]{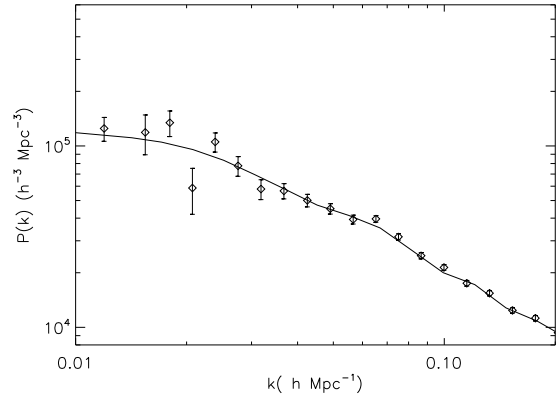}\\
   \end{center}
   \caption{Best fit galaxy power spectrum calculated with the values of table \ref{table1} for the fit to $WMAP7+LRG$ data compared with LRG galaxy power spectrum.}\label{fit}
\end{figure}

\begin{figure*}[ht]
  \begin{center}
    \begin{tabular}{ccc}
        \includegraphics[width=450pt]{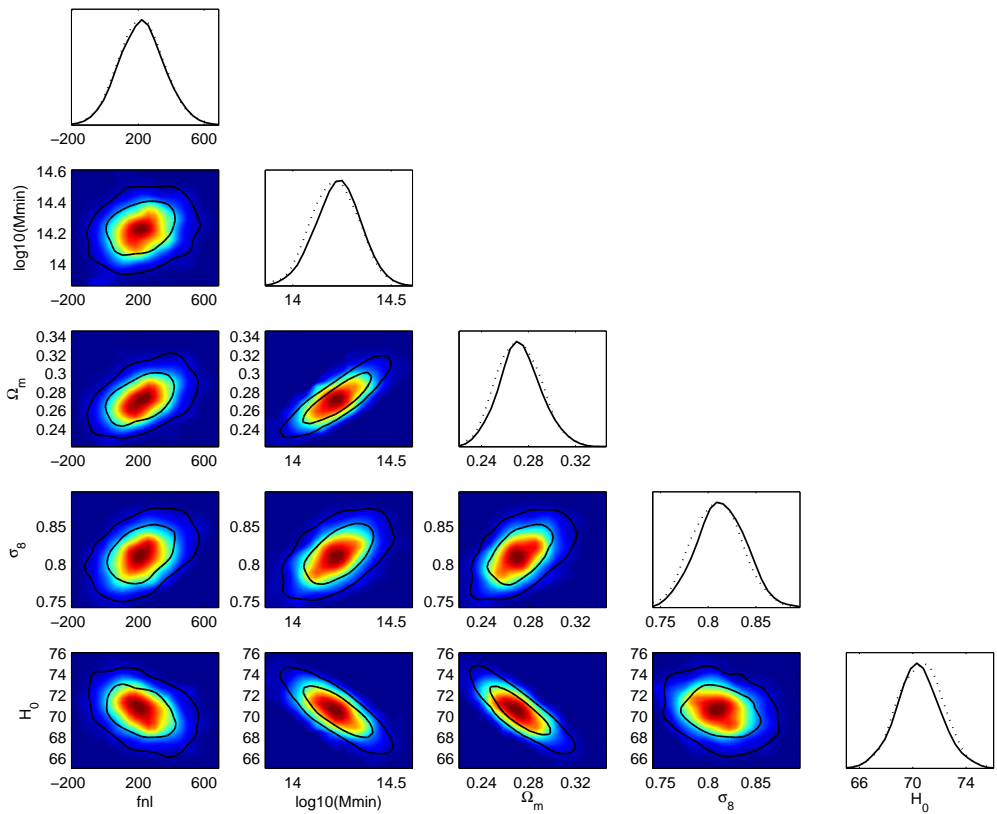}\\
    \end{tabular}
    \caption{$68\%$ and $95\%$ C.L. contour plots and likelihoods for $f_{NL}$ and other parameters of the model for the fit to SDSS DR4 galaxy power spectrum combined with $WMAP7$, $SNe$ and $HST$ data. The plot shows the degeneracies of $f_{NL}$ with cosmological and halo model parameters.
    As expected $f_{NL}$ results to be correlated with matter density and Hubble parameter. Strong degeneracies involve also $M_{min}$, $\Omega_m$, $\sigma_8$ and $H_0$
    weakening the constraints on these parameters.}
    \label{degeneracies}
  \end{center}
\end{figure*}

\begin{table}
\begin{center}
\begin{tabular}{lr|rr}
&$WMAP7+LRG$& & $WMAP7+LRG$\\
&&&$+HST+SNe$\\
\hline \hline
 $ 10^2\Omega_b h^2$        & $2.241^{+0.065}_{-0.063}$      & \hspace{21pt} &  $2.263^{+0.055}_{-0.054} $ \\
 $ \Omega_c h^2$            & $0.1103^{+0.0047}_{-0.0047}$   &  \hspace{21pt}&  $0.1123^{+0.0036}_{-0.0035}  $  \\
 $ \theta$                  & $0.010395^{+0.000032}_{-0.000030}$& \hspace{21pt}&  $0.010396^{+0.000028}_{-0.000027}  $ \\
 $ \tau$                    & $0.088^{+0.0075}_{-0.0087}$      & \hspace{21pt} & $0.087^{+0.0065}_{-0.0072}  $ \\
 $ n_s$                     & $0.964^{+0.014}_{-0.015}$      &  \hspace{21pt} & $0.965^{+0.012}_{-0.013}  $\\
 $ln(10^{10} A_s)$          & $3.08^{+0.04}_{-0.03}$         &  \hspace{21pt} & $3.08^{+0.03}_{-0.03}  $ \\
 $h$                        & $0.715^{+0.023}_{-0.021}$      & \hspace{21pt}     & $0.705^{+0.016}_{-0.015}  $\\
 $\sigma_8$                 & $0.800^{+0.028}_{-0.029}$      & \hspace{21pt} & $0.812^{+0.025}_{-0.025}  $\\
 $\log(M_{min})$             &$13.90^{+0.16}_{-0.25}$        & \hspace{21pt} & $14.19^{+0.12}_{-0.12}  $\\
$\alpha$                     &$0.85^{+0.18}_{-0.20}$         & \hspace{21pt} & $0.83^{+0.17}_{-0.18}  $\\
$\gamma$                          &$9.97^{+5.5}_{-5.6}$           & \hspace{21pt} & $10.7^{+5.1}_{-5.1}  $\\
$\sigma_{cen}$               &$1.00^{+0.57}_{-0.57}$         & \hspace{21pt} & $1.09^{+0.56}_{-0.56}  $ \\
$\log(M_{1})$                &$12.0^{+2.7}_{-2.6}$           & \hspace{21pt} &   $12.3^{+2.7}_{-2.6}  $\\
$f_{NL}$                     &$171^{+140}_{-139}$            & \hspace{21pt} &   $202^{+129}_{-130}  $ \\
 \hline
\hline
\end{tabular}
\caption{Best fit values and $68\%$ C.L. errors on the parameters of our model for $WMAP7+LRG$ and $WMAP7+LRG+SNe+HST$ data. The combination with $SNe$ ans $HST$ data improves only slightly the constraints.}\label{table1}\vspace{1cm}
\end{center}
\end{table}

that reduces to:
\begin{equation}\label{correction}
    \Delta b(M,z,k)=\frac{3\Omega_m
    H_0^2}{c^2k^2T(k)G(z)}f_{NL}(b-r)\delta_{sc}
\end{equation}
where $G(z)$ is the linear growth factor, $T(k)$ is the transfer
function and the parameter $r$ is $1$ if the objects equally
populate all the halos, that is a good assumption for the LRG
galaxies we use in this analysis (see also \cite{Slosar:2008hx}) or
$\sim1.6$ for objects populating only recently merged halos. The
information on $f_{NL}$ is then expected to come from the low-$k$
part of galaxy power spectrum because of the $k^{-2}$ term of the
(\ref{correction}) ($T(k)$ is constant at low wave vectors). For an
even quite large value of $f_{NL}$, for example $f_{NL}=+100$, the
correction on the halo bias is smaller than a $10\%$ for wave
vectors $k>0.01 h Mpc^{-1}$ (see Fig. \ref{biascorrection}). The
amplitude of the correction is proportional to $f_{NL}$ but also to
$H_0$ and $\Omega_m$. One should expect hence important degeneracies
among these parameters that will affect the strength of constraints
on $f_{NL}$.
\begin{figure*}[htbb]
  \begin{center}
    \begin{tabular}{ccc}
  \includegraphics[width=550pt]{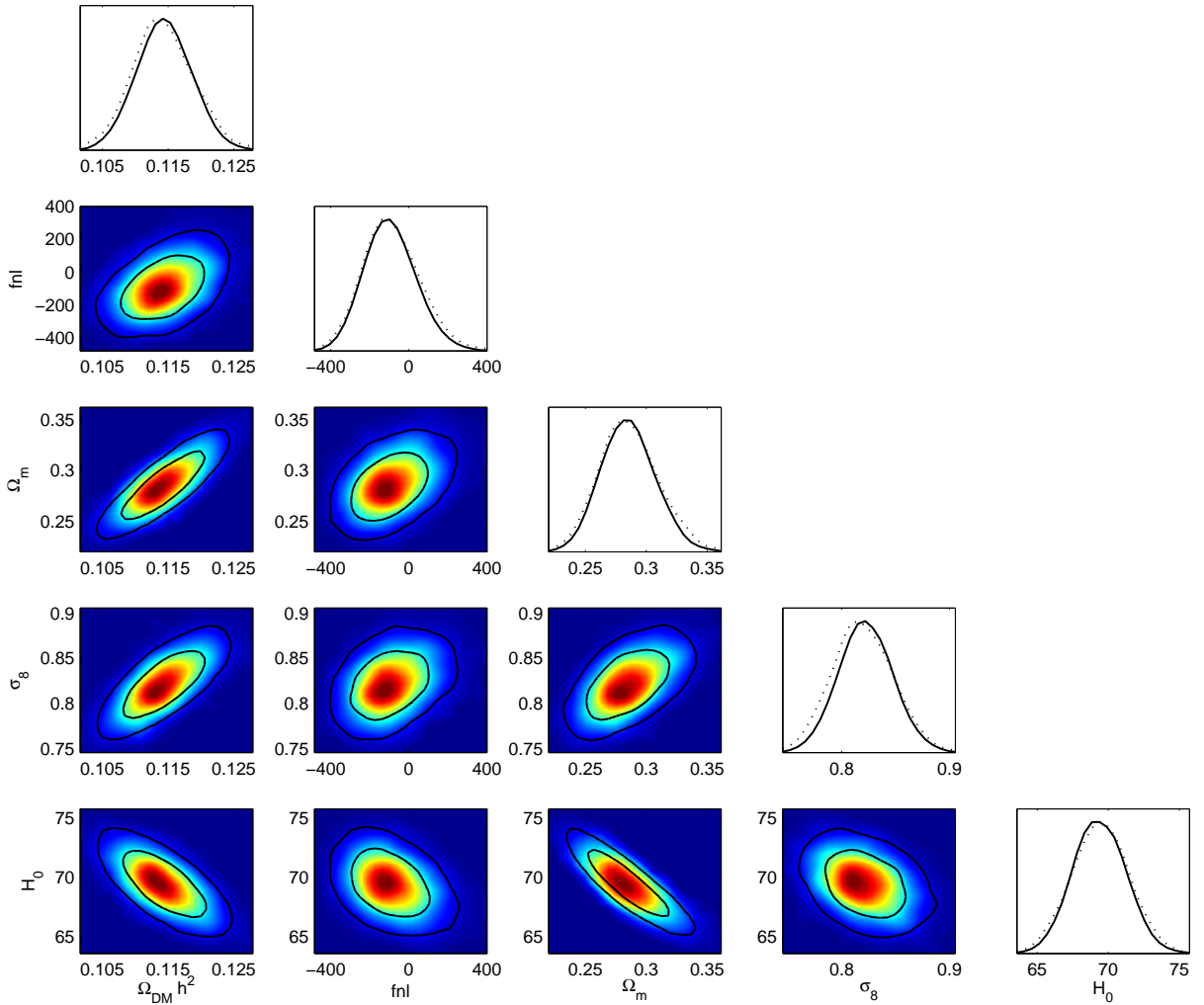}\\
    \end{tabular}
    \caption{$68\%$ and $95\%$ C.L. contour plots and likelihoods for some parameters of our model for the fit to SDSS-DR7 data. As noted before $f_{NL}$ results to be correlated mainly with matter density and Hubble parameter.}
    \label{deg_new}
  \end{center}
\end{figure*}

\section{Analysis and Results}\label{results}
\subsection{Constraints from Red Luminous Galaxies power
spectra}\label{galaxydata} We implemented the calculation of the
theoretical galaxy power spectrum through the halo model described
above and performed a Monte Carlo Markov Chain analysis using Cosmic
Microwave Background data from WMAP-$7$ years of observations
\cite{Komatsu:2010fb} and the most recent LRG galaxy power spectrum
data \cite{Tegmark:2006az} available from Sloan digital Sky Survey
at a mean redshift $z\simeq0.35$. We fitted these data assuming
flatness of the Universe over a $13$ parameters model that consists
of $7$ standard cosmological parameters (the physical baryon and
cold dark matter densities, the ratio of sound horizon to the
angular diameter distance at decoupling, , the optical depth to
reionization, the scalar spectral index, the overall normalization
of the spectrum at $k = 0.002 h Mpc^{-1}$ and the amplitude of
Sunyaev-Zel'dovich spectrum: $\Omega_bh^2$, $\Omega_ch^2$, $\theta$,
$\tau$, $n_s$, $\log_{10}10^{10}A_s$, $A_{SZ}$ ) and of the $5$
parameters of the halo-model plus the non-gaussianity parameter
($\log_{10}M_{min}$, $\alpha$, $\gamma$, $\sigma_{cen}$,
$\log_{10}M_1$, $f_{NL}$). In what follows we will express the
masses in units of solar masses. The Markov Chain analysis has been
performed using the publicly available cosmological code cosmoMC
\cite{Lewis:2002ah} suitably modified to include the calculation of
the halo model and to fit over the parameters of the halo model and
$f_{NL}$. The convergence diagnostic of this code is based on the
Gelman and Rubin statistic \cite{GR} (also known as $R-1$ statistic,
where $R$ is defined as the ratio between the variance of chain
means and the mean of variances). The results of our fit are shown
in table \ref{table1} and Fig. \ref{degeneracies}. For this model we
found weak constraints on the non-gaussianity,
$f_{NL}=171^{+140}_{-139}$ at $68\%$ C.L. and a range
$-69<f_{NL}<+492$ at $95\%$ C.L. from the combination $WMAP7$ and
$LRG$ galaxy power spectrum.\\ The best fit power spectrum computed
for the values of table \ref{table1} is shown in Fig. \ref{fit}; as
one can see there is a slight preference for a non-zero value of
$f_{NL}$ at $1\sigma$ but is largely consistent with gaussian
initial conditions when we consider $2\sigma$ constraints. These
limits are weaker than those obtained with a similar dataset in
\cite{Slosar:2008hx} where small scale non linearities are modeled
with a two parameter $k$ dependent correction. The difference is
that in our case the uncertainty on $f_{NL}$ is heavily affected by
degeneracies with $\Omega_m$, $H_0$ and $\sigma_8$, parameters which
are themselves degenerate with the parameters of the halo model. We
are in fact requiring the information on $f_{NL}$ to come only from
$SDSS$ data since we are using WMAP data only to constrain
cosmological parameters. The $LRG$ data range only for scales
between $0.01<k<0.2 h Mpc^{-1}$ and, as shown in Figure \ref{fit},
on these scales the effect of an even large non-gaussianity is small
and can be easily confused with the effect of cosmological or
halo-model parameters. We repeated this fit including both the
Hubble Space Telescope prior on $H_0$ from \cite{Riess:2009pu} and
Supernovae distance moduli measurements for the Union dataset
\cite{Kowalski:2008ez} obtaining only a slightly improved constraint
on $f_{NL}$, i.e. $-35<f_{NL}<+479$ at $95\%$, and other parameters.
In Fig \ref{degeneracies} we plot constraints on $f_{NL}$ and on the
parameters most involved in degeneracies with $f_{NL}$ for the fit
to $WMAP7+LRG+SNe+HST$ data. As for the model parameters, we found
generally higher values for $\log_{10}M_{min}$, $\gamma$, and
$\sigma_{cen}$ than \cite{Abazajian:2004tn}, but with greater
uncertainty, while $\alpha$ results to be in good agreement and
$\log_{10}M_{1}$ has a very large uncertainty. These differences may
arise because of the different dataset and modeling (in
\cite{Abazajian:2004tn} they fit the projected correlation function)
and parameter space. The same figure confirms the expected
degeneracy between $\Omega_m$ and $log_{10}M_{min}$ as found also in
\cite{Abazajian:2004tn}, due to the correlation between halo masses
and number of galaxies in massive halos \cite{Zheng:2002es},
\cite{Rozo:2004xh} . The consequence of this degeneracy is that our
constraint on the matter density is $\Omega_m=0.264\pm0.022$ from
$WMAP7+LRG$ and hence only slightly better than the constrain from
$WMAP7$ alone $\Omega_m=0.266\pm0.029$.

\subsection{Constraints from halo power spectra}\label{halodata} In this section we
constrain $f_{NL}$ using recent data of power spectrum for the
reconstructed halo density field derived from a sample of LRGs
\cite{Reid:2009xm} in the seventh data release of the SDSS (DR7).
The halo power spectrum is more directly connected to dark matter
density field for a wider $k$ range and this allows to use data
points in the power spectrum up to $k\sim0.2 h Mpc^{-1}$.
\begin{figure}[t]
  \begin{center}
  \includegraphics[width=250pt]{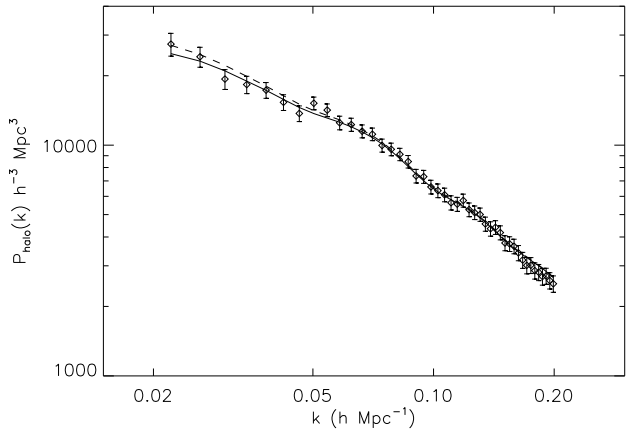}\\
  \end{center}
   \caption{Best fit model (solid line) to halo power spectra data from SDSS-DR7. We show for comparison the
   same model but with $f_{NL}=0$. A negative value of $f_{NL}$ allows to have a better fit to the data points in the range
    $(0.03<k<0.05)hMpc^{-1}$}\label{fitnew}
\end{figure}
 \begin{table}[h]
\begin{center}
\begin{tabular}{lr}
\hline \hline
 $ 10^2\Omega_b h^2$        & $2.248^{+0.055}_{-0.055}$   \\
 $ \Omega_c h^2$            & $0.1144^{+0.0041}_{-0.0041}$ \\
 $ \theta$                  & $0.010389^{+0.000026}_{-0.000027}$   \\
 $ \tau$                    & $0.086^{+0.0063}_{-0.0072}$       \\
 $ n_s$                     & $0.963^{+0.013}_{-0.013}$       \\
 $ln(10^{10} A_s)$          & $3.09^{+0.03}_{-0.03}$         \\
 $h$                        & $0.694^{+0.018}_{-0.018}$ \\
 $\sigma_8$                 & $0.822^{+0.025}_{-0.024}$  \\
 $f_{NL}$                     &$-93^{+128}_{-129}$\\
 \hline
\hline
\end{tabular}
\caption{Best fit values and $68\%$ C.L. errors on the parameters of
our model for the fit to SDSS-DR7 halo power
spectra.}\label{table2}\vspace{1cm}
\end{center}
\end{table}
The main difference with respect to the analysis of the previous
subsection is that to model the halo power spectrum it is not
necessary to model the halo occupation distribution of galaxies. The
halo power spectrum in \cite{Reid:2009xm} is modeled as:
\begin{equation}\label{phalo}
P_{halo}(k)=P_{damp}(k)r_{DM,damp}(k)r_{halo}(k)F_n(k)
\end{equation}

where $P_{damp}$ is a power spectrum that account for damping of
Baryonic Acoustic Oscillations and is calculated as:
\begin{equation}
P_{damp}(k)=P_{0}(k)e^{-\frac{k^2\sigma^2}{2}}+P_{nw}(k)\left(1-e^{-\frac{k^2\sigma^2}{2}}\right)
\end{equation}
with $P_0$ being the linear matter power spectrum and $P_{nw}$ is
the matter power spectrum with baryon oscillations removed
calculated as in \cite{Eisenstein:1997ik}. The value of $\sigma$ is
chosen fitting the reconstructed halo density field in the mock LRG
catalogues \cite{Reid:2009xm}\cite{Reid:2008zu}. The factor $F_n(k)$
is a nuisance term defined as:
\begin{equation}
F_n(k)=b_0^2\left(1+a_1\left(\frac{k}{k_*}\right)+a_2\left(\frac{k}{k_*}\right)^2\right)
\end{equation}
where $b_0$ is the effective bias of the LRG at the effective
redshift $z_{eff}=0.313$ and $k_*=0.2hMpc^{-2}$. The likelihood code
for halo power spectra froe SDSS-DR7 is implemented in cosmoMC and
performs a marginalization over the nuisance parameters $b_0$, $a_1$
and $a_2$. The terms $r_{DM,damp}(k)$ and $r_{halo}(k)$ in (\ref{phalo}) model the
connection between the non-linear matter
power spectrum and the damped linear power spectrum and between halo
and matter power spectrum and they are calibrated against numerical
simulations (see section $3$ in \cite{Reid:2009xm} for more
details). Here we use a modified version of the modeling described
so far introducing the $k$ dependent bias correction
(\ref{correction}) averaged over masses:
\begin{equation}\label{biasave}
b(k,z)=\frac{\int[b(M,z)+\Delta b(M,z,k)]n_{halo}(M,z)MdM}{\int
n_{halo}(M,z)MdM}
\end{equation}
We then fitted the data varying $f_{NL}$ together with the seven
cosmological parameters ($\Omega_bh^2$, $\Omega_ch^2$, $\theta$,
$\tau$, $n_s$, $\log_{10}10^{10}A_s$, $A_{SZ}$) and minimizing
the chi-square by varying nuisance parameters $a_1$ and $a_2$. Our results
are shown in table \ref{table2} and in Figure \ref{deg_new}. We also
show our best fit halo power spectra in Figure \ref{fitnew}. Our
fit for this dataset shows a preference for a negative value
$f_{NL}=-93\pm128$ at $68\%$ C.L. that allows to have a better fit
to five point in the observed power spectra in the range
$(0.03<k<0.05) hMpc^{-1}$. The uncertainty on this value remains
anyway quite large and the $95\%$ C.L. range for $f_{NL}$ is
$-327<f_{NL}<+177$, hence with a very slight improvement with
respect to the constraints from previous dataset. For the other
cosmological parameters we find a good agreement with results from
\cite{Komatsu:2010fb} for WMAP7 combined with halo power spectra of LRG sample.\\ We note that, according
to degeneracies showed in Figure \ref{deg_new}, allowing for a possible non-gaussianity implies an increase
of uncertainty on some cosmological parameters, namely $h$ and $\sigma_8$ with an increase of a $\sim10\%$ on the $1\sigma$ error with respect to $\Lambda CDM$ case for WMAP7+LRG and an increase of a $\sim14\%$ on the error for  $\Omega_ch^2$.
\begin{table}[t]
\begin{center}
\begin{tabular}{lrr}
 & EUCLID+Planck & \\
\hline
Fiducial value&\hspace{50pt}&$\sigma$\\
 \hline \hline
 $f_{NL}=+1$     &  \hspace{50pt}         & $2.23$ \\
 $f_{NL}=+5$     &   \hspace{50pt}        & $2.29$ \\
 $f_{NL}=+10$    &   \hspace{50pt}        & $2.39$ \\
\hline \hline
\end{tabular}
\caption{$1\sigma$ errors ($68\%$C.L.)from the combination of mock
datasets generated for the specifications of Planck experiment and EUCLID survey amd for three different fiducial values of $f_{NL}$.}\label{tabfor}
\end{center}
\end{table}

\section{Forecast for future surveys}\label{forecasts}
In this section we consider constraints on $f_{NL}$ from future data
generating mock datasets for both CMB anisotropy and galaxy power spectra.
For a galaxy
surveys the error on the matter power spectrum can be calculated as
\cite{Feldman}\cite{Tegmark:1997rp}:
\begin{equation}\label{relerr}
    \left(\frac{\sigma_P}{P}\right)^2=\frac{2\pi^2}{4k^2 \Delta kV_{eff}}
\end{equation}
where the effective volume of the survey is given by:
\begin{equation}\label{effvol}
    V_{eff}=V\left(\frac{nP}{nP+1}\right)^2
\end{equation}
and $\Delta k$ is the width of $k$-bins. Here we use specification
for a typical future galaxy survey like EUCLID
\cite{Refregier:2006vt} with galaxy number density $n\simeq1.6\cdot
10^{-3}$, redshift range $0<z<2$ and $f_{sky}\simeq0.5$. The minimum
$k$ of the mock dataset is choose to be greater than $2\pi/V^{1/3}$,
while the maximum $k$ we use is $0.02h Mpc^{-1}$. For CMB anisotropy
power spectrum we use specification for Planck experiment
\cite{:2006uk} assuming the noise of the $143GHz$ channel. We
explore a $\Lambda$ CDM$+f_{NL}$ model and we choose as fiducial
model the WMAP-7 years best fit for $\Lambda$ CDM parameters
\cite{Komatsu:2010fb}. For the non-Gaussianity parameter we choose
three fiducial models, $f_{NL}=+1,+5,+10$. Remember that in our
approach we are using only the information of large scale galaxy
clustering to constrain non-gaussianity, while we use CMB
measurements only to constrain other cosmological parameters and hence to break degeneracies.\\
Results for our forecast on $f_{NL}$ are shown in table
\ref{tabfor}. As one can see the combination of accurate galaxy
power spectrum measurements, attainable with a survey like EUCLID
and Planck CMB measurements could reach the sensitivity required to
detect an even small non-Gaussianity, such as $f_{NL}=+5$ or
$f_{NL}=+10$, with a confidence level of at least $95\%$. We note
that this results is in agreement with other forecast for future
galaxy surveys (see \cite{Carbone:2008iz} for example). Very small
non-Gaussianities ($f_{NL}=+1$) seem instead rather difficult to
detect, mainly due to degeneracies with other parameters.
Nevertheless in \cite{Carbone:2010sb} it has also been shown that in
more complicated models (allowing variation of neutrino mass,
running of spectral index, dark energy equation of state and
relativistic degrees of freedom) constraints on $f_{NL}$ may
deteriorate up to $\sim80\%$.

\section{Systematics}
Before concluding we discuss the possible systematics introduced by the
assumptions we made or, more generally, by the theoretical
uncertainties of the model.\\First, we have seen that the value of
$r$ that appear in the (\ref{correction}) may have a value in the
range $1-1.6$. We have assumed $r=1$ since we are using Luminous Red
Galaxies that are old galaxies at the center of halos. This is a
common assumption for this kind of analysis (see also
\cite{Slosar:2008hx}). Anyway we find that even assuming $r=1.6$ the
differences in the power spectrum with respect to the case $r=1$ are
very small. Using $r=1.6$ we find only a slight variation in the
$\chi^2$ with respect to $r=1$ for the best fit model: $\Delta
\chi^2\sim 0.3$. The reason for this is that actual data constrain
scales $k>0.02 hMpc^{-1}$ where the exact value of $r$ is less
relevant. Nevertheless for future low-$k$ data it may be necessary a
more precise modeling of the (\ref{correction}).\\A second
assumption we made is that the density profile of the halos is
described by the NFW profile. Although the exact shape of the
profiles in the halos is still uncertain, this profile has been
tested against several numerical simulations and it turned out to be
a good approximation \cite{Cooray:2002dia}. Moreover, again, the
information on $f_{NL}$ is coming from $k<0.1 hMpc^{-1}$ where the
density profile is constant in Fourier space.\\An important point is
the comparison between the results from the LRG power spectrum of
\cite{Tegmark:2006az} and the halo power spectrum from
\cite{Reid:2009xm}. The first dataset provides a galaxy power
spectrum while the second a halo density field that doesn't require
any assumption about the halo occupation number. In
\cite{Reid:2009xm} there is a large discussion on the differences
between these two dataset and we refer the reader to this work for a
complete discussion. Here we remark that the main differences are
due to the heavy Finger of Gods compression algorithm used in
\cite{Tegmark:2006az} to obtain the matter power spectrum. This
process may cause transfer of power from a scale to another, causing
consistent deviations (up to $\sim40\%$ on $k=0.2 hMpc^{-1}$)
between the reconstructed halo density field and the matter power
spectrum \cite{Reid:2009xm}. For the DR7 halo power spectrum
instead, the halo density field is reconstructed before the
computation of the power spectrum (see section 2.2 in
\cite{Reid:2009xm}) and the deviations between the two are smaller
than $4\%$. Also the modeling of the theoretical halo power spectrum
and of the galaxy power spectrum is different. The model of
\cite{Reid:2009xm} is calibrated on N-body simulations and mock
data-sets made especially for this LRG sample. Moreover the authors
of \cite{Reid:2009xm} imposed priors on the nuisance parameters of
the model based on N-body simulations. For the galaxy power spectrum
of \cite{Tegmark:2006az} it was used the $Q$-model
\cite{Cole:2005sx} for the non linear part of the power spectrum,
marginalizing over $Q$ with weak priors. All these differences
necessarily reflects on the cosmological parameters estimation,
including $f_{NL}$. The comparison between results of the two
datasets  is made in section 6.1 of \cite{Reid:2009xm}. Significant
differences are found on some cosmological parameters from the two
LRG datasets only (i.e. not including CMB) of the two releases: in
particular the $\Omega_mh$ values (that enters also in the
(\ref{correction})) differ of almost 2$\sigma$ between the two
surveys.\\In our work, to fit the DR7 data we are only introducing
the bias scale dependent correction to the model of
\cite{Reid:2009xm}, in order to be as much as possible consistent
with the data compression algorithm of this LRG catalogue. We
ascribe the differences between the results of section
\ref{galaxydata} and \ref{halodata} to the significant differences
of the data compression process, as noted also in
\cite{Reid:2009xm}.\\A last issue concern the modeling of power
spectra on non-linear scales. For DR7 data, the authors of
\cite{Reid:2009xm} normalize the final halo power spectrum using
mock catalogues to account for the small offset between the N-body
and HALOFIT results.\\Concerning the galaxy power spectrum we used
to fit data from \cite{Tegmark:2006az}, the $P(k)$ we are using in
relation (\ref{2halo}) is the linear matter power spectrum, which is
well known. The galaxy power spectrum is calculated through the
halo-model itself. The $5$-parameters model we are using showed a
good agreement with hydrodynamical and N-body simulations and
semi-analytic models \cite{Guzik:2002zp}, \cite{Berlind:2002rn},
\cite{Kravtsov:2003sg}, \cite{Viero:2009qm}, \cite{Berlind:2001xk}.
Moreover, we are marginalizing over the $5$ free parameters that
account for the uncertainties of the model, so that our analysis is
rather conservative.\\In the forecast section we model the galaxy
power spectrum relying on the same assumption made above ($r=1$ for
galaxies and NFW profiles) and using the same HOD modeling. Our
results show the potential of a survey like Euclid to detect even
small non gaussianities and are in good agreement with forecast done
for the same survey and for a similar modeling of the scale
dependent bias \cite{Carbone:2008iz}. It is clear, however that the
analysis of real data from these future surveys will probably
require a more accurate modeling of the galaxy power spectrum and of
the scale dependent correction in order to not bias the estimated
value of the cosmological parameters.

\section{Conclusions}
We place new constraints on the local type non-gaussianity parameter
$f_{NL}$ by looking at the scale dependence of the halo bias (at
small wave vectors) in the recent galaxy and halo power spectra
measurements from LRG sample of the Sloan Digital Sky Survey. We fit
2006 SDSS power spectra data with an halo model consisting of 5
parameters plus 7 cosmological paramers and $f_{NL}$. Our large
parameter space and the the restriction of the dataset to relatively
small scales ($k>0.01 h Mpc^{-1}$) leads to a weak constraint:
$-69<f_{NL}<+492$ at $95\%$ C.L.. We show and discuss degeneracies
with halo model parameters. When including both Type Ia Supernovae
and $HST$ data the $1\sigma$ error on $f_{NL}$ is reduced of about
$\sim10\%$. We use also 2009 halo power spectra data obtained from
SDSS LRG sample finding a slightly better constraint
$-327<f_{NL}<+177$ at $95\%$ C.L., again limited by the fact that
the dataset does not extend below $k\sim0.02h Mpc^{-1}$. We also
forecast the constraints obtainable from datasets of a survey like
EUCLID when combined with Planck CMB data, finding that these
surveys could reach the accuracy required to detect even small
non-gaussianities as $f_{NL}=+5$, thus confirming the power of this
method. Finally we discuss the possible systematics and theoretical
uncertainties that may affect the results.

\textit{Acknowledgements}\\\\
FDB thanks UCI Center for Cosmology for support and hospitality
while this research was conducted. This work was supported by NSF
CAREER AST-0645427 and NASA NNX10AD42G at UCI. This research has
been partially supported by the ASI/INAF agreement I/072/09/0 for
the Planck LFI Activity of Phase E2.


\begin{thebibliography}{99}

\bibitem{starob}
A.~A. {Starobinski{\v i}}, Soviet Journal of Experimental and Theoretical
  Physics Letters {\bf 30},  682  (1979).

\bibitem{guth}
A.~H. {Guth}, \prd {\bf 23},  347  (1981).

\bibitem{linde}
A.~D. {Linde}, Physics Letters B {\bf 108},  389  (1982).

\bibitem{albrecht}
A. {Albrecht} and P.~J. {Steinhardt}, Physical Review Letters {\bf 48},  1220
  (1982).

\bibitem{M}
V. F. Mukhanov and G. V. Chibisov, Soviet Journal of
Experimental and Theoretical Physics Letters 33, 532
(1981).

\bibitem{Hawking1982}
S. W. Hawking, Physics Letters B 115, 295 (1982).

\bibitem{Guth1982}
A. H. Guth and S.-Y. Pi, Physical Review Letters 49,
1110 (1982).

\bibitem{Starobinkij1982}
A. A. Starobinskij, Physics Letters B 117, 175 (1982).

\bibitem{Bardeen1983}
J. M. Bardeen, P. J. Steinhardt, and M. S. Turner, Phys.
Rev. D 28, 679 (1983).


\bibitem{bardeen}
J. M. Bardeen, J. R. Bond, N. Kaiser, and A. S. Szalay, Astrophys. J. 304, 15 (1986).

\bibitem{linde2}
A. Linde and V. Mukhanov, Phys. Rev. D56, R535 (1997), astro-ph/9610219.

\bibitem{lythwands}
D. H. Lyth and D. Wands, Phys. Lett. B524, 5 (2002), hep-ph/0110002.

\bibitem{lythung}
D. H. Lyth, C. Ungarelli, and D. Wands, Phys. Rev. D67, 023503 (2003), astro-ph/0208055.

\bibitem{vernizzi}
F. Vernizzi and D. Wands, JCAP 0605, 019 (2006), astro-ph/0603799.

\bibitem{Malik}
K. A. Malik and D. H. Lyth, JCAP 0609, 008 (2006), astro-ph/0604387.

\bibitem{sasaki}
M. Sasaki, J. Valiviita, and D. Wands, Phys. Rev. D74, 103003 (2006), astro-ph/0607627.

\bibitem{Bartolo:2010qu}
  N.~Bartolo, S.~Matarrese and A.~Riotto,
  arXiv:1001.3957 [astro-ph.CO].

\bibitem{Matarrese:1997sk}
  S.~Matarrese, L.~Verde and A.~F.~Heavens,
  Mon.\ Not.\ Roy.\ Astron.\ Soc.\  {\bf 290} (1997) 651
  (1997MNRAS.290..651M)


\bibitem{lu_mat}
F. Lucchin, S. Matarrese, and N. {Vittorio}, \aap {\bf 162},  13  (1986).

\bibitem{matave:ApJ}
S. {Matarrese}, L. {Verde}, and R. {Jimenez}, \apj {\bf 541},  10  (2000).

\bibitem{Rob:MNRAS}
J. {Robinson} and J.~E. {Baker}, \mnras {\bf 311},  781  (2000).

\bibitem{Benson:MNRAS}
A.~J. {Benson}, C. {Reichardt}, and M. {Kamionkowski}, \mnras {\bf 331},  71
  (2002).

\bibitem{Scoccimarro:2003wn}
  R.~Scoccimarro, E.~Sefusatti and M.~Zaldarriaga,
  Phys.\ Rev.\  D {\bf 69} (2004) 103513
  [arXiv:astro-ph/0312286].

\bibitem{Dalal:2007cu}
  N.~Dalal, O.~Dore, D.~Huterer and A.~Shirokov,
  Phys.\ Rev.\  D {\bf 77} (2008) 123514
  [arXiv:0710.4560 [astro-ph]].

\bibitem{Gangui:1993tt}
  A.~Gangui, F.~Lucchin, S.~Matarrese and S.~Mollerach,
  Astrophys.\ J.\  {\bf 430} (1994) 447
  [arXiv:astro-ph/9312033].

\bibitem{Komatsu:2000vy}
  E.~Komatsu and D.~N.~Spergel,
Published in Proceedings. Edited by R.T. Jantzen, V. Gurzadyan and
R. Ruffini, Singapore, World Scientific, 2002. pp.2009
  arXiv:astro-ph/0012197.


\bibitem{Komatsu:2010fb}
  E.~Komatsu {\it et al.},
  arXiv:1001.4538 [astro-ph.CO].

\bibitem{Komatsu:2008hk}
  E.~Komatsu {\it et al.}  [WMAP Collaboration],
  Astrophys.\ J.\ Suppl.\  {\bf 180} (2009) 330
  [arXiv:0803.0547 [astro-ph]].

\bibitem{Calabrese:2009bu}
  E.~Calabrese {\it et al.},
  Phys.\ Rev.\  D {\bf 81} (2010) 043529
  [arXiv:0909.1837 [astro-ph.CO]].

\bibitem{Rudjord:2009mh}
  O.~Rudjord, F.~K.~Hansen, X.~Lan, M.~Liguori, D.~Marinucci and S.~Matarrese,
  Astrophys.\ J.\  {\bf 701} (2009) 369
  [arXiv:0901.3154 [astro-ph.CO]].

\bibitem{Verde:1999ij}
  L.~Verde, L.~M.~Wang, A.~Heavens and M.~Kamionkowski,
  Mon.\ Not.\ Roy.\ Astron.\ Soc.\  {\bf 313} (2000) L141
  [arXiv:astro-ph/9906301].

\bibitem{Reid:2010vc}
  B.~A.~Reid, L.~Verde, K.~Dolag, S.~Matarrese and L.~Moscardini,
  arXiv:1004.1637 [astro-ph.CO].


\bibitem{Matarrese:2008nc}
  S.~Matarrese and L.~Verde,
  Astrophys.\ J.\  {\bf 677} (2008) L77
  [arXiv:0801.4826 [astro-ph]].

\bibitem{Slosar:2008hx}
  A.~Slosar, C.~Hirata, U.~Seljak, S.~Ho and N.~Padmanabhan,
  JCAP {\bf 0808} (2008) 031
  [arXiv:0805.3580 [astro-ph]].

\bibitem{Xia:2010yu}
  J.~Q.~Xia, M.~Viel, C.~Baccigalupi, G.~De Zotti, S.~Matarrese and L.~Verde,
  arXiv:1003.3451 [astro-ph.CO].


\bibitem{Percival:2009xn}
  W.~J.~Percival {\it et al.},
  Mon.\ Not.\ Roy.\ Astron.\ Soc.\  {\bf 401} (2010) 2148
  [arXiv:0907.1660 [astro-ph.CO]].

\bibitem{Kowalski:2008ez}
  M.~Kowalski {\it et al.}  [Supernova Cosmology Project Collaboration],
  Astrophys.\ J.\  {\bf 686} (2008) 749
  [arXiv:0804.4142 [astro-ph]].

\bibitem{condon98}
Condon, J. J., Cotton, W. D., Greisen, E. W., Yin, Q. F., Perley,
R. A., Taylor, G. B., Broderick, J. J. 1998, Astron. J., 115,
1693

\bibitem{Cooray:2002dia}
  A.~Cooray, R.~K.~Sheth,
  Phys.\ Rept.\  {\bf 372} (2002) 1
  [arXiv:astro-ph/0206508].


\bibitem{www.sdss.org}
www.sdss.org

\bibitem{Tegmark:2006az}
  M.~Tegmark {\it et al.}  [SDSS Collaboration],
  Phys.\ Rev.\  D {\bf 74} (2006) 123507
  [arXiv:astro-ph/0608632].

\bibitem{Reid:2009xm}
  B.~A.~Reid {\it et al.},
  arXiv:0907.1659 [astro-ph.CO].



\bibitem{Riess:2009pu}
  A.~G.~Riess {\it et al.},
  Astrophys.\ J.\  {\bf 699} (2009) 539
  [arXiv:0905.0695 [astro-ph.CO]].


\bibitem{Refregier:2006vt}
  A.~Refregier {\it et al.},
  Proc. SPIE, Vol. 6265, 62651Y (2006)
  arXiv:astro-ph/0610062.

\bibitem{:2006uk}
  G.Efstathiou, C.Lawrence, J.Tauber and the Planck Collaboration,
  arXiv:astro-ph/0604069.


\bibitem{whiterees}
  White, S.D.M., Rees, M. 1978, Mon.\ Not.\ Roy.\ Astron.\ Soc.\ textbf{183}
  341
  Phys.\ Rept.\  {\bf 372} (2002) 1
  [arXiv:astro-ph/0206508].


\bibitem{Peacock:2000qk}
  J.~A.~Peacock and R.~E.~Smith,
  Mon.\ Not.\ Roy.\ Astron.\ Soc.\  {\bf 318} (2000) 1144
  [arXiv:astro-ph/0005010].

\bibitem{Sheth:2001dp}
  R.~K.~Sheth and G.~Tormen,
  Mon.\ Not.\ Roy.\ Astron.\ Soc.\  {\bf 329} (2002) 61
  [arXiv:astro-ph/0105113].

\bibitem{Abazajian:2004tn}
  K.~Abazajian {\it et al.}  [SDSS Collaboration],
  Astrophys.\ J.\  {\bf 625} (2005) 613
  [arXiv:astro-ph/0408003].

\bibitem{Zheng:2004id}
  Z.~Zheng {\it et al.},
  Astrophys.\ J.\  {\bf 633} (2005) 791
  [arXiv:astro-ph/0408564].

\bibitem{Guzik:2002zp}
  J.~Guzik and U.~Seljak,
  Mon.\ Not.\ Roy.\ Astron.\ Soc.\  {\bf 335} (2002) 311
  [arXiv:astro-ph/0201448].

\bibitem{Berlind:2002rn}
  A.~A.~Berlind {\it et al.},
  Astrophys.\ J.\  {\bf 593} (2003) 1
  [arXiv:astro-ph/0212357].


\bibitem{Kravtsov:2003sg}
  A.~V.~Kravtsov, A.~A.~Berlind, R.~H.~Wechsler, A.~A.~Klypin, S.~Gottloeber, B.~Allgood and J.~R.~Primack,
  Astrophys.\ J.\  {\bf 609} (2004) 35
  [arXiv:astro-ph/0308519].

\bibitem{Navarro:1996gj}
  J.~F.~Navarro, C.~S.~Frenk and S.~D.~M.~White,
  Astrophys.\ J.\  {\bf 490} (1997) 493
  [arXiv:astro-ph/9611107].

\bibitem{Magliocchetti:2006qa}
  M.~Magliocchetti, L.~Silva, A.~Lapi, G.~De Zotti, G.~L.~Granato, D.~Fadda and L.~Danese,
  Mon.\ Not.\ Roy.\ Astron.\ Soc.\  {\bf 375} (2007) 1121
  [arXiv:astro-ph/0611409].

\bibitem{Viero:2009qm}
  M.~P.~Viero {\it et al.},
  Astrophys.\ J.\  {\bf 707} (2009) 1766
  [arXiv:0904.1200 [astro-ph.CO]].

\bibitem{Berlind:2001xk}
  A.~A.~Berlind and D.~H.~Weinberg,
  Astrophys.\ J.\  {\bf 575} (2002) 587
  [arXiv:astro-ph/0109001].

\bibitem{Press&Schechter}
Press, W.H., Schechter, P. 1974, ApJ, \textbf{187}, 425

\bibitem{Sheth:1999mn}
  R.~K.~Sheth and G.~Tormen,
  Mon.\ Not.\ Roy.\ Astron.\ Soc.\  {\bf 308} (1999) 119
  [arXiv:astro-ph/9901122].

\bibitem{Mo:1996zb}
  H.~J.~Mo, Y.~P.~Jing and S.~D.~M.~White,
  Mon.\ Not.\ Roy.\ Astron.\ Soc.\  {\bf 282} (1996) 1096
  (1996MNRAS.282.1096M)

\bibitem{Komatsu:2001rj}
  E.~Komatsu and D.~N.~Spergel,
  Phys.\ Rev.\  D {\bf 63} (2001) 063002
  [arXiv:astro-ph/0005036].




\bibitem{Lewis:2002ah}
  A.~Lewis and S.~Bridle,
  Phys.\ Rev.\  D {\bf 66} (2002) 103511
  [arXiv:astro-ph/0205436].

\bibitem{GR}
A. Gelman and D. B. Rubin., Statist.\ Sci.\ Vol.\ 7, Number 4
(1992), 457-472

\bibitem{Zheng:2002es}
  Z.~Zheng, J.~L.~Tinker, D.~H.~Weinberg and A.~A.~Berlind,
  Astrophys.\ J.\  {\bf 575} (2002) 617
  [arXiv:astro-ph/0202358].

\bibitem{Rozo:2004xh}
  E.~Rozo, S.~Dodelson and J.~A.~Frieman,
  Phys.\ Rev.\  D {\bf 70} (2004) 083008
  [arXiv:astro-ph/0401578].


\bibitem{Eisenstein:1997ik}
  D.~J.~Eisenstein and W.~Hu,
  Astrophys.\ J.\  {\bf 496} (1998) 605
  [arXiv:astro-ph/9709112].

\bibitem{Reid:2008zu}
  B.~A.~Reid, D.~N.~Spergel and P.~Bode,
  Astrophys.\ J.\  {\bf 702} (2009) 249
  [arXiv:0811.1025 [astro-ph]].

\bibitem{Feldman}
Feldman, H. A., Kaiser, N., Peacock, J. A. 1994, ApJ, 426, 23

\bibitem{Tegmark:1997rp}
  M.~Tegmark,
  Phys.\ Rev.\ Lett.\  {\bf 79} (1997) 3806
  [arXiv:astro-ph/9706198].


\bibitem{Carbone:2008iz}
  C.~Carbone, L.~Verde and S.~Matarrese,
  Astrophys.\ J.\  {\bf 684} (2008) L1
  [arXiv:0806.1950 [astro-ph]].

\bibitem{Carbone:2010sb}
  C.~Carbone, O.~Mena and L.~Verde,
  arXiv:1003.0456 [astro-ph.CO].


\bibitem{Cole:2005sx}
  S.~Cole {\it et al.}  [The 2dFGRS Collaboration],
  Mon.\ Not.\ Roy.\ Astron.\ Soc.\  {\bf 362} (2005) 505
  [arXiv:astro-ph/0501174].
\end{thebibliography}
\end{document}